\documentstyle[twoside,fleqn,espcrc2]{article}

\newcommand{\AmS}{{\protect\the\textfont2
  A\kern-.1667em\lower.5ex\hbox{M}\kern-.125emS}}

\input epsf

\title{Quenched QED on a momentum space lattice
\thanks{Work supported in part by the National Science Fundation under 
grant NSF-PHY92-00148. The numerical simulations were performed on the 
CRAY C-90's at PSC and NERSC.}}

\author{J.B. Kogut \address{Department of Physics, University of Illinois, 
        1110 West Green Street, Urbana IL 61801-3080, USA} 
        and 
        J.-F. Laga\"e \address{Department of Physics, University of Kentucky,
        395 Chem.-Phys. Bldg., Lexington KY 40506, USA}}
       
\begin{document}

\begin{abstract}
We discuss the advantages of using the momentum space lattice method for 
studying the phase diagram of quenched $\mbox{QED}_4$. Preliminary results 
of a numerical simulation are presented. They indicate that the method avoids 
the contamination by 4-fermi interactions which plagues the conventional 
position space non-compact formulation.
\end{abstract}

\maketitle

At last year's conference, we introduced the momentum space lattice method
and discussed its general features \cite{Lagae94}. 
Here we would like to report on its
application to quenched $\mbox{QED}_4$. Before going into the details of
the simulation, however, it is worth explaining the advantages of the 
method for the problem at hand. We will therefore start with a brief reminder
of earlier analytical and numerical results and motivate from there the 
introduction of the momentum space idea.

The analytical work of Miransky, Bardeen, Leung and Love (MBLL) 
\cite{MBLL} indicates that
the phase diagram of quenched QED should be described in the ($\alpha$,$G$)
plane, where $\alpha$ is the usual QED coupling constant and $G$ is the 
strength of a 4-fermi interaction. From the ladder Schwinger-Dyson equation,
it is in fact possible to determine the shape of the critical line and to 
estimate the value of the critical exponents along this line \cite{Kocic90}: 
They vary
continuously from their mean-field value in the pure 4-fermi theory to an
essential singularity at the MBLL point. Numerical simulations of quenched
non-compact $\mbox{QED}_4$ appear to pick-up a particular point along 
this line ($\alpha=0.44\alpha_c$, $G=3.06$), 
implying that non-compact QED actually
has to be interpreted as a ``mixture'' of QED + 4-fermi. The values of the
critical exponents at this point ($\delta=2.2(1)$, $\beta=0.78(8)$, 
$\nu=0.68(2)$) were confirmed in more extensive simulations \cite{Kocic93}  
and recent results strengthen the hypothesis of power-law scaling 
\cite{Kocic94}. 
All of this however 
raises an important question: can we find a simulation technique that would 
allow us to move along the critical line and expose the physics of ``pure'' 
QED? We certainly expect that if such a technique can be found in the quenched 
case, it will play an important role in future simulations of the full theory.

A key point for carrying-out the above program would be to identify precisely 
the origin of 4-fermi interactions in non-compact QED with action:
\begin{eqnarray}
\hspace{-0.7cm}&&S = S_G + S_F 
= {\beta \over 4} \sum_{x,\mu,\nu}(\triangle_\mu \theta_\nu)^2 \\
\hspace{-0.7cm}&&
-{1 \over 2} \sum_{x,\mu} \eta_{\mu}(x)\bar{\psi}_x [ U_\mu(x) \psi_{x+\mu}
        -U^*_\mu(x-\mu) \psi_{x-\mu} ] \nonumber
\end{eqnarray} 
An interesting candidate is revealed by using the quadratic nature of $S_G$ 
and computing exactly 
the integration over the gauge fields. This operation leads to an effective 
action of the form:
\begin{eqnarray}
S_{\mbox{eff.}} & = & {1 \over 4} \sum_{x,\mu} (\bar{\psi}\psi)_x
                      (\bar{\psi}\psi)_{x+\mu} \nonumber \\
 & + & \exp[-V(0)/2\beta] S_F(U=1) \\
 & + & \cdots \vphantom{\sum}\nonumber
\end{eqnarray}
where we have assumed
\begin{equation}
D_{\mu\nu}(x-y)=\delta_{\mu\nu}V(x-y)
\end{equation}
The first term represents a 4-fermi interaction, the second is a tadpole 
improved kinetic energy term, and the dots stand for the current-current 
interaction as well as 3-body and higher order interactions. The form of (2) 
suggests that in order to get an effective action which is closer to 
``pure'' QED, we should try to get rid of the 4-fermi term. In fact,
recent results on quenched models seem to indicate that this term, when 
present, is very efficient in masking the rest of the dynamics: critical 
exponents appear to be largely insensitive to the choice of the potential
V \cite{Hands}. 
There is no objection in principle to subtracting the 4-fermi term 
directly from the original action. It is gauge invariant and has all the 
appropriate symmetries (in fact, it is just the strong coupling limit of  
(1)). However, in practice, this turns out to be rather 
difficult. One could think for example of subtracting the 4-fermi interaction 
(multiplied by $\lambda^2$) by adding to the action:
\begin{equation}
S_{+}={\lambda \over 2} \sum_{x,\mu} \bar{\psi}_x
  [ e^{i\phi_\mu(x)} \psi_{x+\mu} + e^{-i\phi_\mu(x-\mu)} \psi_{x-\mu} ]
\end{equation}
where $\phi_\mu$ is a new vector field which is freely integrated over 
between 0 and $2\pi$. Notice the relative + sign in (4) opposite to (1).
This technique is
however very slow numerically. In many respects, the problems are analogous 
to those encountered in simulations at finite chemical potential, where a 
term similar to (4) would appear at first order in an expansion of the 
lagrangian in powers of $\mu$ (in this analogy, $\lambda$ is identified with 
the chemical potential $\mu$). As in the chemical potential case, only small 
values of $\lambda$ can be considered here, so that the 4-fermi interaction 
could only be subtracted partially. A more radical modification of the model 
therefore seems necessary.

An important aspect of the 4-fermi interaction in $S_{\mbox{eff.}}$ 
is that it is directly 
related to the compact nature of the coupling of the fermions to the gauge 
field in the original model (as can be seen through the steps leading to (2)).
On the other hand, the momentum space lattice method \cite{Lagae94} 
has a {\it non-compact} 
coupling and therefore avoids the generation of 4-fermi interactions. In 
fact, the method simply uses the continuum lagrangian of QED.
The model is investigated on a torus of length $L$ and the momentum expansion 
is truncated.
After gauge fixing (to the Feynman gauge) and Fourier expansion, the action 
reads:
\begin{eqnarray}
S &=& -{1 \over 2}\beta\sum_q \theta_{\mu}^*(q)(2\pi q)^2 \theta_{\mu}(q)
\nonumber \\
  &+& \sum_k \overline{\chi}(k)[ \gamma_{\mu}(2 \pi k_{\mu}) + \rho ] \chi (k)
\\
  &-& \sum_{k,k^\prime} \overline{\chi}(k) \gamma_{\mu} \theta_{\mu}(k-k^\prime)
      \chi (k^\prime) \nonumber
\end{eqnarray}
where, instead of the usual fields $\psi$ and $A_\mu$, we have used the
following dimensionless variables:
\begin{equation}
 \chi(k)=L^{5/2} \psi(k) \hskip 1cm
   \theta_{\mu}(q) = eL^{3}A_{\mu}(q)
\end{equation}
and $ \beta={1/{e^2}} \hskip .20cm$,
$\hskip .20cm \rho=mL $.
In the above formula, the $k_\mu{}'s$ are integers or half-integers depending
on whether the boundary conditions for the fermions are periodic or 
antiperiodic. The $q_\mu{}'s$ are integers and take all the possible values
of the momentum transfer between $k$ and $k^\prime$. The gauge fields therefore
live on a lattice which is twice as big (in each direction) as the fermionic
lattice. The method has several advantages among which is the fact that it does
not suffer from the fermion doubling problem. From the numerical point of view,
the procedure is however more involved than conventional position space lattice
methods, the reason being the non-locality of the coupling of the fermions to
the gauge field. This is partially taken care of by repeated use of a fast
Fourier transform which reduces the computer time requirements to O($N \log N$).
When quoting our results below, 
we give the size of the ``photon lattice'' since it is the one used in the
FFT and therefore determines the computer time requirements. This size in turn
is constrained by the specifics of the FFT routine and the choice of the
boundary conditions for the fermions. Up to now, we have considered lattices
of size $(6)^4$, $(10)^4$ and $(15)^4$.
We have measured the mass gap and $<\bar{\psi}\psi>$ by respectively 
using point and volume sources on the momentum space lattice (both methods 
make use of the fact that the propagator should be diagonal in momentum after 
statistical average).
Identifying the mass gap M with the inverse of the correlation length,
we expect  at the critical coupling:
\begin{equation}
M \propto m^{\nu/\beta\delta}
\end{equation}
From our $(10)^4$ measurements , we obtain the estimates $\beta_c=0.8$
and $\beta\delta/\nu\simeq2$. We can go one step further and write the
equation of state in the form:
\begin{equation}
{ m \over M^{\beta\delta/\nu} } = F\left( {\triangle\beta \over M^{1/\nu} }
\right) \hspace{1.0cm}  \triangle\beta=\beta-\beta_c
\end{equation}
Fixing $\beta\delta/\nu=2.$ and varying $\beta_c$ and $\nu$, we obtain a best
fit to our full data sample for $\beta_c=0.79$ and $\nu=1.22$.
This value of $\nu$ is anomalously large and indicates that we are indeed
probing the physics of a non-trivial theory (By comparison, the value of $\nu$
obtained on a position space lattice was $\nu=0.68$ and it would be $\nu=0.5$
in mean field theory). Similarly, the value $\beta\delta/\nu\simeq2$ is
consistent with Miransky scaling. It is important to complete this picture
by independent measurements of  the critical indices $\beta$ and $\delta$. 
We measured $<\bar{\psi}\psi>$ on lattices of size $(10)^4$ and $(15)^4$ 
and found respectively $\delta\simeq2.0$ and $\delta\simeq1.7$. The trend 
here is interesting and indicates that as the size of the lattice is increased, 
it becomes possible to detect exponents which are further away from their 
mean-field value (see fig. 1).
\begin{figure}[t]
\vspace{-1.2cm}
\epsfxsize=7.5cm
\epsfbox{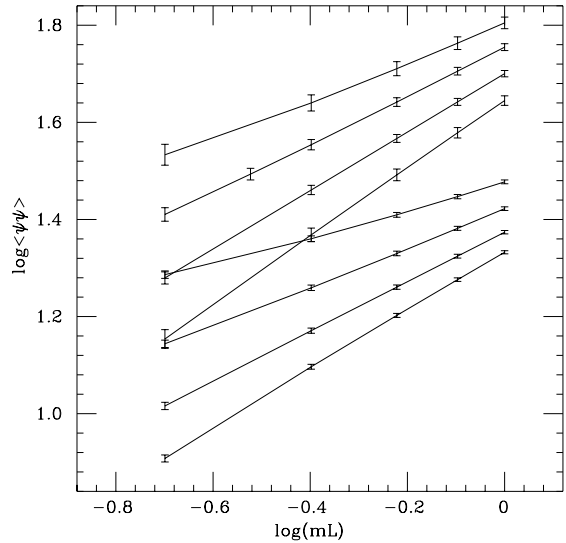}
\vspace{-2cm}
\caption{ Chiral condensate vs bare quark mass at various values of the 
coupling in the vicinity of the transition on a $(10)^4$ lattice (lower group) 
and on a $(15)^4$ lattice (upper group)}
\end{figure}
Since, on a position space lattice, distinguishing between the critical 
behaviour of QED and a Nambu Jona-Lasinio model is a difficult task, it 
is interesting to see how the momentum space lattice behaves in this 
respect. For that purpose, we ran simulations in which we replaced 
$(2 \pi q)^2$ in equation (5) by a constant, so that the long range QED 
potential becomes a point-like interaction.
The estimates for $\delta$ are now $\delta=3.5\pm0.2$ on a $(10)^4$ lattice 
and $\delta=3.2\pm0.1$ on a $(15)^4$ lattice. These can very easily be 
discriminated from the QED values given above and we reach the conclusion 
that on a momentum space lattice, the shape of the potential really matters
 (at the difference of the position space situation \cite{Hands}).

The results presented above are obviously very preliminary and more detailed 
analyses are clearly needed. However, they all indicate that the momentum 
space lattice method is successful in removing 4-fermi interactions from 
QED simulations. The trends that we have established as a function of the 
lattice size, open the way for more detailed simulations of QED on larger 
lattices and the extraction of ``infinite volume'' critical exponents. 
The method of course also has its limitations: it is computationally very 
expensive due to the non-locality of the interaction and a careful analysis 
of its gauge invariance properties is needed. It remains however, 
so far, the only method we have with the potential to map out the entire 
critical line of the quenched model.

\end{document}